\documentclass [prl,superscriptaddress,showpacs,twocolumn]{revtex4}
\usepackage {graphicx}

\begin {document}

\title {Coulomb Parameter $U$ and Correlation Strength in LaFeAsO}

\author {V.~I.~Anisimov}
\affiliation {Institute of Metal Physics, Russian Academy of Sciences,
620041 Yekaterinburg GSP-170, Russia}

\author {Dm.~M.~Korotin}
\affiliation {Institute of Metal Physics, Russian Academy of Sciences,
620041 Yekaterinburg GSP-170, Russia}

\author {S.~V.~Streltsov}
\affiliation {Institute of Metal Physics, Russian Academy of Sciences,
620041 Yekaterinburg GSP-170, Russia}

\author {A.~V.~Kozhevnikov} 
\affiliation {Institute of Metal Physics,
Russian Academy of Sciences, 620041 Yekaterinburg GSP-170, Russia}
\affiliation {Joint Institute for Computational Sciences, Oak Ridge
National Laboratory P.O. Box 2008 Oak Ridge, TN 37831-6173, USA}

\author {J.~Kune\v{s}}
\affiliation {Theoretical Physics III, Center for Electronic Correlations
and Magnetism, Institute of Physics, University of Augsburg, Augsburg
86135, Germany}

\author {A.~O.~Shorikov}
\affiliation {Institute of Metal Physics, Russian Academy of Sciences,
620041 Yekaterinburg GSP-170, Russia}

\author {M.~A.~Korotin}
\affiliation {Institute of Metal Physics, Russian Academy of Sciences,
620041 Yekaterinburg GSP-170, Russia}

\begin {abstract}

\textit {First principles} constrained density functional theory scheme in
Wannier functions formalism has been used to calculate Coulomb repulsion
$U$ and Hund's exchange $J$ parameters for iron $3d$ electrons in LaFeAsO.
Results strongly depend on the basis set used in calculations: when O-$2p$,
As-$4p$, and Fe-$3d$ orbitals and corresponding bands are included,
computation results in $U=$3$\div$4~eV, however, with the basis set
restricted to Fe-$3d$ orbitals and bands only, computation gives parameters
corresponding to $F^0$=0.8~eV, $J$=0.5~eV. LDA+DMFT (the Local Density
Approximation combined with the Dynamical Mean-Field Theory) calculation
with this parameters results in weakly correlated electronic structure that
is in agreement with X-ray experimental spectra.

\end {abstract}


\maketitle

Following the discovery of high-$T_c$ superconductivity in iron oxypnictide
LaO$_{1-x}$F$_x$FeAs~\cite {Kamihara-08}, a question of the influence of
electronic correlation effects on the normal and superconducting properties
of LaFeAsO has arisen. In striking similarity with high-$T_c$ cuprates,
undoped material LaFeAsO is not superconducting with antiferromagnetic
commensurate spin density wave developing below 150~K~\cite {neutrons}. 
Only when electrons (or holes) are added to the system via doping,
antiferromagnetism is suppressed and superconductivity appears. As it is
generally accepted that Coulomb correlations between copper $3d$ electrons
are responsible for cuprates anomalous properties, it is tempting to
suggest that the same is true for iron $3d$ electrons in LaFeAsO.

\begin {figure}
\includegraphics [width=0.425\textwidth]{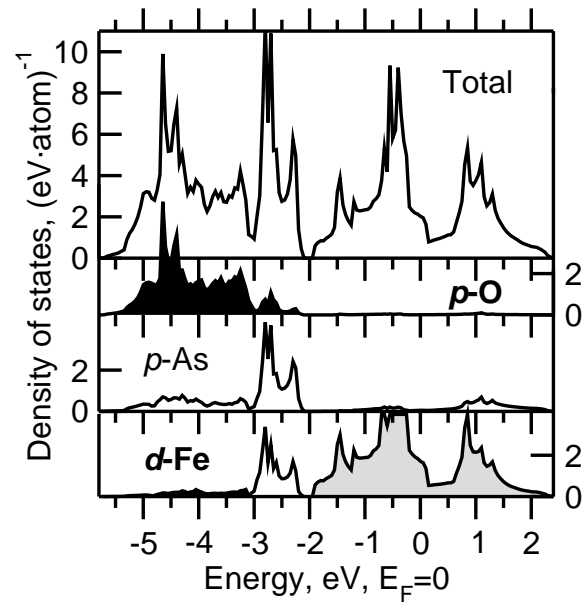}
\caption {Total and partial densities of states for LaFeAsO
obtained in DFT calculation in frame of LMTO method.}
\label {fig1} 
\end {figure}

Correlation strength in a system is determined by ratio of Coulomb
interaction parameter $U$ and band width $W$. If $U/W$ is significantly
less than 1 then the system is weakly correlated and results of the Density
Functional Theory (DFT) calculations are reliable enough to explain its
electronic and magnetic properties. However, if $U$ value is comparable
with $W$ or even larger then the system is in intermediate or strongly
correlated regime and Coulomb interactions must be explicitly treated in
electronic structure calculations. For LaFeAsO the bands formed by Fe-$3d$
states have width $\approx$4~eV (see shaded area in the lower panel of
Fig.~\ref {fig1}), so an estimation for Coulomb interaction parameter
$U$ should be compared with this value.

In practical calculations, $U$ is often considered as a free parameter to
achieve the best agreement of calculated and measured properties of
investigated system. Sometimes $U$ value could be estimated from the
experimental spectra. The \textit {first principles} justified methods to
determine Coulomb interaction parameter $U$ value are constrained DFT
scheme~\cite {U-calc}, where in DFT calculations the $d$-orbital
occupancies are fixed to the certain values and $U$ is numerically
determined as a derivative of $d$-orbital energy over its occupancy and 
Random Phase Approximation (RPA) method, where screened Coulomb interaction
between $d$-electrons is calculated via perturbation theory~\cite {RPA}. In
Ref.~\onlinecite {Haule08} it was proposed to use in LaFeAsO $U$=4~eV
obtained in RPA calculations for metallic iron~\cite {Ferdi}.

This value for Coulomb parameter (with Hund's exchange parameter $J$=0.7 eV)
was used in Dynamical Mean-Field Theory (DMFT)~\cite {DMFT} calculations
for LaFeAsO \cite {Haule08, DMFT-our, Craco08}. Results of these works 
show iron 3$d$ electrons being in intermediate or strongly correlated
regime, as it is natural to be expected for Coulomb parameter value
$U$=4~eV and Fe-$3d$ band width $\approx$4~eV. 

The most direct way to estimate correlation effects strength in a system
under consideration is to compare the experimental spectra with  densities
of states (DOS) obtained in DFT calculations. For strongly correlated
materials additional features in the experimental photoemission and
absorption spectra appear that are interpreted as lower and upper Hubbard
bands absent in the DFT DOS. If no such additional features are observed
and DOS obtained in DFT calculations satisfactorily describe the
experimental spectra then the material is in weakly correlated regime. 

LaFeAsO was studied by soft X-ray absorption and emission
spectroscopy~\cite {X-ray}, X-ray absorption spectroscopy (O
$K$-edge)~\cite {exp-U} and photoemission spectroscopy~\cite {Malaeb-08}.
In all these works the conclusion was that DOS obtained in DFT calculations
gave good agreement with the experimental spectra and the estimations for
Coulomb parameter value are $U<$1 eV~\cite {exp-U}. That contradiction with
results of the DMFT calculations~\cite {Haule08, DMFT-our, Craco08} shows
that \textit {first principles} calculation of  Coulomb interaction
parameter $U$ value for LaFeAsO is needed to determine the correlation
effects strength in this material. Results of such calculations by
constrained DFT calculations are reported in the present work. We have
obtained the value $U<$1~eV for Fe-$3d$ band that agrees with the estimates
from spectroscopy. Recently the RPA calculations for Coulomb interaction
parameter $U$ in LaFeAsO were reported, where $U$ was estimated as
1.8$\div$2.7 eV~\cite {Arita}. 

It is important to note that Coulomb interaction parameter $U$ value
depends on the model where it will be used and, more precisely, on the
choice of the orbital set that is taken explicitly into account in the
model. For example, in constrained DFT calculations for high-$T_c$ cuprates
the resulting $U$ value for Cu $d$-shell was found between 8 and
10~eV~\cite {cuprate-d}. The $U$ value in this range was used in cluster
calculations where all Cu $d$-orbitals and $p$-orbitals of neighboring
oxygens were taken into account and calculated spectra agree well with
experimental data~\cite {cluster-Cu}. However, in one band model, where
only $x^2-y^2$ orbital per cooper atom is explicitly included in the
calculations, the $U$ value giving good agreement with experimental data
falls down to 2.5$\div$3.6~eV~\cite {cuprate-2D}, that is 3-4 times smaller
than constrained DFT value. 

The same situation occurs for titanium and vanadium oxides: the $U$ value
from constrained DFT calculations is $\approx$6~eV and cluster calculations
where all $d$-orbitals and $p$-orbitals of neighboring oxygens were taken
into account with $U$ close to this value gave good agreement between
calculated and experimental spectra~\cite {cluster-Ti-V}. However, in the
model where only partially filled $t_{2g}$ orbitals are included, much
smaller $U$ value (corresponding to Slater integral $F^0$=3.5~eV) gives the
results in agreement with experimental data~\cite {V-Ti-dmft}.

It is interesting that such a small $U$ value can be obtained in
constrained DFT calculations for titanates and vanadates where only
$t_{2g}$-orbital occupancies are fixed while all the other states
($e_g$-orbitals of vanadium and $p$-orbitals of oxygens) allowed to relax
in self-consistent iterations~\cite {vanadates-titanates, V-Ti-dmft}. So
the calculation scheme used in constrained DFT (the set of the orbitals
with fixed occupancies) should be consistent with basis set of the model
where the calculated $U$ value will be used.

Another source of uncertainty in constrained DFT calculation scheme is a
definition of atomic orbitals whose occupancies are fixed and energy
calculated. In some DFT methods, like Linearized Muffin-Tin Orbitals
(LMTO)~\cite {LMTO}, these orbitals could be identified with LMTO. However,
in other DFT calculation schemes, where plane waves are used as a basis,
like in pseudopotential method~\cite {PW} one should use more general
definition for localized atomic like orbitals such as Wannier functions
(WFs)~\cite {Wannier37}. The practical way to calculate WFs for specific
materials using projection of atomic orbitals on Bloch functions was
developed in \cite {MarzariVanderbilt}. 

In Fig.~\ref {fig1} the total and partial DOS for LaFeAsO obtained in 
LMTO calculations are shown. Crystal field splitting for Fe-$3d$ orbitals
in this material is rather weak ($\Delta_{cf}$=0.25~eV) and all five $d$
orbitals of iron form common band in the energy region ($-$2, $+$2)~eV
relative to the Fermi level (see grey region on the bottom panel in
Fig.~\ref {fig1}). There is a strong hybridization of iron $t_{2g}$
orbitals with $p$ orbitals of arsenic atoms which form nearest neighbors
tetrahedron around iron ion. This effect becomes apparent in the energy
interval ($-$3, $-$2)~eV (white region on the bottom panel in Fig.~\ref
{fig1}) where band formed by $p$ orbitals of arsenic is situated. More week
hybridization with oxygen $p$ states reveals in ($-$5.5, $-$3)~eV energy
window (black region on the bottom panel in Fig.~\ref {fig1}).

We have calculated Coulomb interaction $U$ and Hund's exchange $J$
parameters for WFs basis set via constrained DFT procedure with fixed
occupancies for WFs of $d$ symmetry. For this purpose we have used two
calculation schemes based on the different DFT methods. One of them
involves linearized muffin-tin orbitals produced by the TB-LMTO-ASA
code~\cite {LMTO}; corresponding WFs calculation procedure is described in
details in Ref.~\onlinecite {WF-LMTO}. The second one is based on the plane
waves obtained within the pseudopotential plane-wave method PWSCF, as
implemented in the Quantum ESPRESSO package~\cite {PW}, and described in
details in Ref.~\onlinecite {WF-PW}. The difference between the results of
these two schemes could give an estimation for the error of $U$ and $J$
determination.

The WFs are defined by the choice of Bloch functions Hilbert space and by a
set of trial localized orbitals that will be projected on these Bloch
functions. We performed calculations for two different choices of Bloch
functions and atomic orbitals. One of them includes only bands
predominantly formed by Fe-$3d$ orbitals in the energy window ($-$2,
$+$2)~eV and equal number of Fe-$3d$ orbitals to be projected on the Bloch
functions for these bands. That choice corresponds to the model where only
five $d$-orbital per Fe site are included but all arsenic and oxygen
$p$-orbitals are omitted. Second choice includes all bands in energy window
($-$5.5, $+$2)~eV that are formed by O-$2p$, As-$4p$ and Fe-$3d$ states and
correspondingly full set of O-$2p$, As-$4p$ and Fe-$3d$ atomic orbitals to
be projected on Bloch functions for these bands. That would correspond to
the extended model where in addition to $d$-orbitals all $p$-orbitals are
included too.

\begin {figure}
\includegraphics [width=0.425\textwidth]{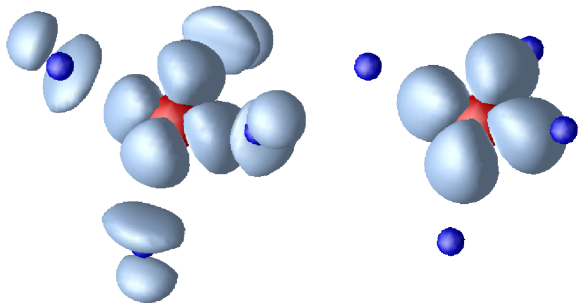}
\caption {(Color online) Module square of $d_{x^2-y^2}$-like Wannier
function computed for Fe-$3d$ bands only (left panel) and for full set of
O-$2p$, As-$4p$ and Fe-$3d$ bands (right panel). Big sphere in the center
marks Fe ion position and four small spheres around it correspond to As
neighbors.}
\label {fig:WF}
\end {figure}

In both cases we obtained Hamiltonian in WF basis that reproduces exactly
bands predominantly formed by Fe-$3d$ states in the energy window ($-$2,
$+$2)~eV (Fig.~\ref {fig1}), but in the second case in addition to that
bands formed by $p$-orbitals in the energy window ($-$5.5, $-$2)~eV will be
reproduced too. However, WFs with $d$-orbital symmetry computed in those
two cases have very different spatial distribution. In Fig.~\ref {fig:WF}
the module square of $d_{x^2-y^2}$-like WF is plotted. While for the case
when full set of bands and atomic orbitals was used (right panel) WF is
nearly pure atomic d-orbital (iron states contribute 99\%), WF computed
using Fe-$3d$ bands only is much more extended in space (left panel). It
has significant weight on neighboring As ions with only 67\% contribution
from central iron atom. 

The physical reason for such effect is $p-d$ hybridization that is treated
explicitly in the case where both $p$- and $d$-orbitals are included. In
the case where only Fe-$3d$ bands are included in calculation $p-d$
hybridization reveals itself in the shape of WF. Fe-$3d$ bands in the
energy window ($-$2, $+$2)~eV correspond to antibonding combination of
Fe-$3d$ and As-$4p$ states and that is clearly seen on the left panel of
Fig.~\ref {fig:WF}.

The different spatial distribution for two WFs calculated with full and
restricted orbital bases can be expected to lead to different effective
Coulomb interaction for electrons occupying these states. The results of
constrained DFT calculations of the average Coulomb interaction $\bar{U}$
and Hund's exchange $J$ parameters for electrons on WFs computed with two
different set of bands and orbitals (and using two different DFT methods:
LMTO and pseudopotential) are presented in Tab.~\ref {tab:UJ}.

\begin {table}
\caption {The constrained DFT calculated values of average Coulomb
interaction $\bar{U}$ and Hund's exchange $J$ (eV) parameters for 
$d$-symmetry Wannier functions computed with two different sets of bands
and orbitals.}
\begin {tabular}{l|c|c}
\hline \multicolumn {1}{l}{DFT method} & \multicolumn {1}{|c}{separate
Fe-$3d$ band} & \multicolumn {1}{|c}{full bands set} \\ \hline TB-LMTO-ASA
&  $\bar{U}$=0.49, $J$=0.51 & $\bar{U}$=3.10, $J$=0.81 \\ PWSCF &
$\bar{U}$=0.59, $J$=0.53 &  $\bar{U}$=4.00, $J$=1.02 \\ \hline 
\end {tabular}
\label {tab:UJ} 
\end {table}

One can see that very different Coulomb interaction strength is obtained
for separate Fe-$3d$ band and full bands set calculations. While the latter
gives value 3$\div$4~eV, that agrees with previously used values~\cite
{Haule08, DMFT-our, Craco08}, separate Fe-$3d$ band calculation results in
0.5$\div$0.6~eV, that is much smaller but agrees with spectroscopy
estimations~\cite {exp-U}.

The main reason for such a drastic difference between two calculations is
very different spatial extension of the two WFs (see Fig.~\ref {fig:WF}):
nearly complete localization on central iron atom for ``full bands set'' WF
(99\%) and only 67\% for ``Fe-$3d$ band set'' WF. Another possible reason
for strong reduction of the calculated $\bar{U}$ value in going from ``full
bands set'' to ``Fe-$3d$ band set'' WF is screening via $p-d$ hybridization
with As-$4p$ band that is situated just below Fe-$3d$ band (see Fig.~\ref
{fig1}). The effect of decreasing of the effective $\bar{U}$ value in
several times going from full orbital model to restricted basis was found
previously for high-$T_c$ cuprates ($U$=8$\div$10~eV for full
$p-d$-orbitals basis~\cite {cuprate-d} and 2.5$\div$3.6~eV for one-band
model~\cite {cuprate-2D}).

In constrained DFT calculations one obtains an average Coulomb interaction
$\bar{U}$ that can be estimated as $\bar{U}=F^0-J/2$. Hence, Slater
integral $F^0$ can be calculated as $F^0=\bar{U}+J/2$~\cite {U-calc}. For
``Fe-$3d$ band set'' WF that gives $F^0$=0.8~eV at $J$=0.5~eV. With this
set of parameters we performed the LDA+DMFT~\cite {LDA+DMFT} calculations
(for detailed description of the present computation scheme see
Ref.~\onlinecite {WF-LMTO}). The DFT band structure was calculated within
the TB-LMTO-ASA method~\cite {LMTO}. Crystal structure parameters were
taken from Ref.~\onlinecite {Kamihara-08}. 

\begin {figure}
\includegraphics [width=0.425\textwidth]{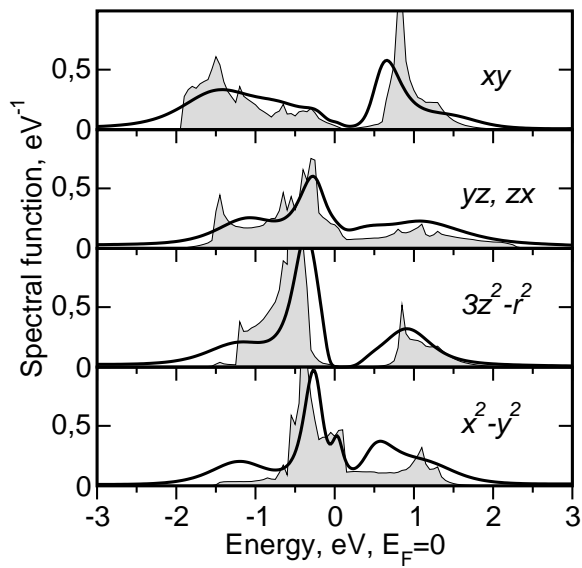}
\caption {Partial densities of states for different Fe-$3d$ orbitals
obtained within the DFT (filled areas) and LDA+DMFT orbitally resolved
spectral functions for $F^0$=0.8~eV, $J$=0.5~eV (bold lines).}
\label {LDA+DMFT-dos} 
\end {figure}

The restricted basis set including only Fe-$3d$ WFs was used in the
LDA+DMFT calculations. The effective impurity model for the DMFT was solved
by the QMC method in Hirsh-Fye algorithm~\cite {HF86}. Calculations were
performed for the value of inverse temperature $\beta$=10~$eV^{-1}$.
Inverse temperature interval $0<\tau<\beta$ was divided into 100 slices. 
$6 \cdot 10^6$ QMC sweeps were used in self-con\-sis\-ten\-cy loop within
the LDA+DMFT scheme and $12 \cdot 10^6$ of QMC sweeps were used to
calculate the spectral functions.

The iron $3d$ orbitally resolved spectral functions obtained within DFT and
LDA+DMFT  calculations are presented in Fig.~\ref {LDA+DMFT-dos}. The
influence of correlation effects on the electronic structure of LaFeAsO is
minimal: there are small changes of peak positions for $3z^2-r^2$, $xy$ and
$x^2-y^2$ orbitals (the shift toward the Fermi energy) and practically
unchanged picture of spectral function distribution for $yz, zx$ bands.
There are no appearance of neither quasiparticle peak on the Fermi level
nor Hubbard bands in the energy spectrum with such values of $U$ and $J$.
Hence LaFeAsO can be considered as weakly correlated material.

This agrees with the results of soft X-ray absorption and emission
spectroscopy study~\cite {X-ray}. It was concluded there that LaFeAsO does
not represent strongly correlated system since Fe $L_3$ X-ray emission
spectra do not show any features that would indicate the presence of the
low Hubbard band or the quasiparticle peak that were predicted by the
LDA+DMFT analysis~\cite {Haule08, DMFT-our, Craco08} with the large
$U$=4~eV. A comparison of the X-ray absorption spectra (O $K$-edge) with
the LDA calculations gave an upper limit of the on-site Hubbard
$U\approx$1~eV~\cite {exp-U}. Photoemission spectroscopy study of LaFeAsO
suggests~\cite {Malaeb-08} that the line shapes of Fe $2p$ core-level
spectra correspond to an itinerant character of Fe $3d$ electrons. It was
demonstrated there that the valence-band spectra are generally consistent
with band-structure calculations except for the shifts of Fe $3d$-derived
peaks toward the Fermi level. 

In conclusion, we have calculated the values of $U$ and $J$ via constrained
DFT procedure in the basis of WFs. For minimal model including only Fe-$3d$
orbitals we have obtained Coulomb parameters $F^0$=0.8~eV, $J$=0.5~eV.
The LDA+DMFT calculation with these parameters results in weakly correlated
nature of iron $d$ bands in this compound. This conclusion is supported by
several X-ray spectroscopic investigations of this material.

Support by the Russian Foundation for Basic Research under Grant No.
RFFI-07-02-00041, Civil Research and Development Foundation together with
Russian Ministry of science and education through program Y4-P-05-15,
Russian president grant for young scientists MK-1184.2007.2 and Dynasty
Foundation is gratefully acknowledged. J.K. acknowledges the support of SFB
484 of the Deutsche Forschungsgemeinschaft.

\end {document}